\documentclass[12pt]{article}
\usepackage{latexsym}
\usepackage{amsfonts}
\newcommand{\R}{\ensuremath{\textrm{I}\!\textrm{R}}}

\newcommand{\g}{\ensuremath{\mathfrak{g}}}
\newcommand{\V}{\ensuremath{\mathcal{V}}}
\newcommand{\h}{\ensuremath{\mathfrak{h}}}
\newcommand{\ka}{\ensuremath{\mathfrak{k}}}
\newcommand{\eq}{\begin{equation}}
\newcommand{\en}{\end{equation}}
\newcommand{\nn}{\nonumber}
\newcommand{\eqn}{\begin{eqnarray}}
\newcommand{\enn}{\end{eqnarray}}
\newcommand{\eqns}{\begin{eqnarray*}}
\newcommand{\enns}{\end{eqnarray*}}
\newcommand{\dm}{\begin{displaymath}}
\newcommand{\dn}{\end{displaymath}}
\begin{document}
\begin{titlepage}
\begin{flushright}
  PSU-TH-223\\
\end{flushright}
\begin{center}
\begin{LARGE}
\textbf{Dimensionally Reduced Gravity, Hermitian Symmetric Spaces and the 
Ashtekar Variables}
\end{LARGE}\\
\vspace{1cm}
Othmar Brodbeck\\
\vspace{.25cm}
\emph{Max-Planck-Institute for Physics\\
Werner Heisenberg Institute\\ 
D-80805 Munich, Germany}\\
\vspace{.75cm}
Marco Zagermann\footnote{zagerman@phys.psu.edu}  \\
\vspace{.25cm}
\emph{Department of Physics\\
Pennsylvania State University\\
104 Davey Lab\\
University Park, PA 16802, USA} \\
\vspace{1cm}
{\bf Abstract} 
\end{center}
\begin{small}

Dimensional reductions of various higher dimensional (super)gravity 
theories lead to 
effectively two-dimensional field theories described by gravity coupled 
$G/H$ nonlinear $\sigma$-models. We show that a new set of complexified 
variables 
can be introduced when  $G/H$ is a  
Hermitian symmetric space.
This generalizes an earlier  construction that grew out of the Ashtekar 
formulation of two Killing vector reduced  pure $4d$ general relativity. 
Apart from giving 
some new insights into dimensional reductions of higher dimensional
(super)gravity theories, these Ashtekar-type variables
offer several technical advantages in the context of the exact 
quantization of these models. As an application, an infinite 
set of conserved charges is constructed. Our results might serve as a 
starting point for probing the \emph{quantum} equivalence of the 
Ashtekar and the metric formalism within a non-trivial midi-superspace model 
of 
quantum gravity.

\end{small}

\end{titlepage}

\renewcommand{\theequation}{\arabic{section}.\arabic{equation}}
\section{Introduction}
\setcounter{equation}{0}

Dimensional reductions of higher dimensional (super)gravity theories to two
dimensions have been studied from various points of view. In the physically 
most interesting cases, these theories can be described by an effectively 
$2d$ field theory consisting of a $G/H$ nonlinear 
$\sigma$-model coupled to $2d$ gravity and a dilaton.

From a particle physicist's perspective, these models provide 
somewhat extreme
examples of so-called `hidden' symmetries, as they typically arise 
 in dimensionally reduced (super)gravity theories.
Whereas such `hidden' symmetry groups are finite dimensional 
for reductions to $d\geq 3$ spacetime dimensions, they inflate to  infinite
dimensional symmetry groups in two dimensions.
This phenomenon was first encountered by Geroch \cite{RG}
in the two Killing vector reduction of source-free
$4d$ general relativity, where
$G/H=SL(2,\R)/SO(2)$, and was later found to be a generic 
feature  of the analogous models with more complicated 
coset  spaces $G/H$ that descend from other higher dimensional 
(super)gravity theories \cite{BJ}.

From a general relativist's point of view, these  models, 
although being essential truncations of higher dimensional gravity theories,
still exhibit a surviving $2d$ diffeomorphism invariance
and involve self-interacting \emph{local} degrees of freedom. This  raises 
 the hope
that their exact quantization might give  insights    into 
at least some of the problems of quantum gravity.

What makes such a quantization seem feasible, on the other hand, is 
precisely the  aforementioned rich symmetry structure of these midi-superspace 
models.
On quite general 
grounds (see eg. \cite{HN}), one expects  the 
underlying generators of the Geroch group  (resp. its 
generalizations) to provide an infinite
set of nontrivial observables whose Poisson algebra 
should translate into a spectrum generating algebra in the corresponding 
quantum 
theory. 

In order to make this idea more 
explicit, two    main routes have been pursued.
One approach is to use the formulation of these 
theories in terms of the conventional  metric variables and  to exploit the
existence of the linear system (`Lax pair') \cite{DM,BZ} that encodes the 
metric-based field equations. This was the  strategy of
refs. \cite{KS1,KS2,KS3}, where a Poisson (and eventually  quantum) 
realization of the Geroch group could be achieved for various sets of boundary
conditions.

An alternative approach is offered 
by the Hamiltonian formulation 
of $4d$ general relativity in terms of the connection-type variables put 
forward by Ashtekar \cite{AA}. This direction has been followed
in \cite{HS,VH1,VH2,GM}, where the two Killing vector reduction of
pure $4d$ general relativity has been performed  within
the Ashtekar formulation.

It is the purpose of this paper to demonstrate that this alternative
approach in terms of the Ashtekar variables provides  some  fruitful
insights into both the dimensional reductions of higher dimensional
(super)gravity theories and the non-perturbative 
quantization of these midi-superspace models
in the context of canonical quantum gravity.

Our starting point is the two Killing vector reduction of pure $4d$ 
 Einstein gravity in terms of the Ashtekar variables. As was  shown
in \cite{VH1,MZ}, this reduction naturally 
leads to an interesting set
of $2d$ variables which circumvents several technical difficulties
associated with the traditional metric variables. Most importantly, 
their Poisson brackets are completely ultralocal 
(ie. they don't contain derivatives of delta-functions), which is in contrast
to the $\sigma$-model currents in the conventional metric formulation,
whose non-ultralocal Poisson    brackets  require careful treatment
in the canonical formalism \cite{KS1,KS2,KS3}.   

From a purely
$2d$ point of view, the existence of these alternative variables,
and the simplifications they provide, can be traced back to some
very special properties of the underlying coset space $SL(2,\R)/SO(2)$ 
\cite{MZ}. 

In the first part of this paper, we identify these `very special'
properties as those of Hermitian symmetric spaces (ie., spaces of the form
$G/H=G/(H'\times U(1))$). This implies that all higher dimensional
(super)gravity theories which lead to  Hermitian symmetric space
non-linear $\sigma$-models in two dimensions admit  the construction 
of these $2d$  Ashtekar-type variables. This suggests an interesting interplay 
between the Ashtekar formulation, group theory and the dimensional reduction
of higher dimensional (super)gravity theories.
In fact, our results are consistent with a recent observation made in
\cite{CJLP}, where it was pointed out that all $2d$ Hermitian symmetric 
space $\sigma$-models 
have their ``genuine'' origin in four spacetime dimensions,
which is also the critical dimension for the existence of the Ashtekar 
variables.

In the second part of this paper, we will then have a closer look at the 
technical advantages of  these Ashtekar-type variables  in the 
context 
the canonical quantization of these models. In particular, we will
show that the corresponding field equations also admit the formulation 
in terms of a linear 
system. In contrast to the analogous
linear system for the metric variables \cite{DM,BZ}, 
however, this linear system
can be written in terms of
 a \emph{constant} (ie. spacetime \emph{in}dependent) spectral parameter 
without 
that unwieldy square roots have to be introduced into the linear system.

As an application of this technical simplification, which seems to be  
closely related to the ultralocality of the 
corresponding Poisson structure, we construct an infinite set of conserved 
charges and determine their Poisson brackets by using techniques similar to 
those
in \cite{KS1,KS2,KS3}, which in our case, however,  simplify considerably.

Apart from offering a complementary approach towards a systematic
 quantization of an 
important midi-superspace model,  
the results of this second part can also be used as a starting point for 
probing the \emph{quantum} equivalence of the Ashtekar and the metric
formulation within a non-trivial toy model of quantum gravity,
since the use of similar techniques in both approaches might facilitate
a comparison of the resulting quantum theories.

The organization of the paper is as follows:
For convenience of the reader, we briefly recall, in section 2, 
the results of the two Killing vector reduction of pure 
$4d$ general relativity    in terms of metric 
and Ashtekar variables and  show how the resulting $2d$ formulations 
are related to each other. 
In section 3 we will embed the (metric-based) two Killing vector reduction 
of general 
relativity into the more general class of $2d$ gravity coupled
$G/H$ nonlinear $\sigma$-models as they arise in dimensional reductions of 
various (super)gravity models. This section will mainly serve to establish
our notation. Section 4 then introduces the alternative variables as they 
grew out of the Ashtekar formulation and generalizes their 
construction to arbitrary  Hermitian symmetric coset spaces. The linear system
encoding the field equations of these alternative variables will be 
displayed in Section 5, which also contains a comparison with the 
Lax pair for the metric variables and a construction of an infinite set of 
conserved non-local charges.
 Section 6  concludes with a short discussion of our results.

\section{Motivation: Two Killing vector reduced pure 
$4d$ gravity}\label{motivation}\setcounter{equation}{0}
Before considering more general dimensionally reduced gravity models,
and in order to motivate the more abstract constructions in the rest of
this paper, let us briefly recall the two Killing vector reduction of 
source-free 
$4d$ general  relativity in terms of metric and Ashtekar variables (for
details, see \cite{HS, VH1, GM, MZ}).

Assuming the existence of two commuting, spacelike and 2-surface 
orthogonal\footnote{By 2-surface orthogonality we mean that the two-dimensional
planes orthogonal to the planes spanned by the two Killing vectors are 
integrable. In contrast to \emph{hyper}surface orthogonality, this is a very 
weak (in fact, often even redundant) assumption, which removes some 
topological degrees of freedom right from the beginning.} Killing vector 
fields, one may choose local coordinates 
$(t,x,y,z)\equiv (x^{0}, x^{1}, x^{2}, x^{3})$ such that the Killing vectors 
are given by 
$\frac{\partial}{\partial y}$ and $\frac{\partial}{\partial z} $ and the  
metric  $G_{MN}$  attains the following form
\begin{equation}\label{sum0}
G_{MN} = G_{MN}(t,x)  =
\left(\begin{array}{cc}
e^{2\sigma}\eta_{\mu\nu} & 0\\
0 &\rho  g_{\bar{m}\bar{n}}
\end{array}\right),
\end{equation}
where $M,N,\ldots \in  \{t,x,y,z\};\quad$  $\mu,\nu,\ldots \in  \{t,x\};\quad$
$\bar{m},\bar{n},\ldots \in  \{y,z\}\quad$;
$\det g_{\bar{m}\bar{n}}   = 1$; 
$\eta_{\mu\nu} = \textrm{diag}\,(-1,+1)$ and $\sigma$ and $\rho$ are  
some functions of $(t,x)$.
Rewriting the  $(2\times 2)$-matrix $g$ in terms of \emph{unimodular} 
`zweibeins' $e_{\bar{m}}^{\bar{a}}\in SL(2,\R)$ ($\bar{a}, \bar{b},
\ldots \in \{ 2,3\}$)
\begin{equation}\label{ge}
g_{\bar{m}\bar{n}}= e_{\bar{m}}^{\bar{a}}\delta_{\bar{a}\bar{b}}
e_{\bar{n}}^{\bar{b}},
\end{equation}
it becomes clear that $g$ parametrizes the coset space 
$SL(2,\R)/SO(2)$.

The Einstein equations imply  
$\eta^{\mu\nu}\partial_{\mu}\partial_{\nu}\rho =0$ and a first order equation 
for the conformal factor $\sigma$. The non-trivial dynamics is captured 
by the field $g$. Its field equation reads
\begin{equation}\label{sum1}
\partial_{0}J_{0}-\partial_{1}J_{1}=0.
\end{equation}
where the currents $J_{\mu}$ are defined by
\begin{equation}\label{Jg}
J_{\mu}:=\rho g^{-1}\partial_{\mu}g,
\end{equation}
and therefore obey the following compatibility condition 
\begin{equation}\label{sum2}
\partial_{0}J_{1}-\partial_{1}J_{0}+\frac{1}{\rho}[J_{0},J_{1}]-
\frac{\partial_{0}\rho}{\rho}J_{1}+\frac{\partial_{1}\rho}{\rho}J_{0}=
0.
\end{equation}
The underlying symmetry-reduced $2d$ Lagrangian induces Poisson brackets 
between the currents $J_{\mu}$, which contain non-ultralocal 
terms (see section 3.2). 

In terms of the Ashtekar formulation, on the other hand, 
the symmetry reduction  looks    as follows.

Starting point is the parametrization of the phase space of 
$4d$ general relativity in terms of  the inverse dreibein density 
$\tilde{E}_{a}^{m}$ and the 
Ashtekar connection
$A_{nb}$, where $a,b, \ldots =1,2,3$ are the internal SO(3)-indices,
whereas $m,n,\ldots=x,y,z$. 

$\tilde{E}_{a}^{m}$ and $A_{nb}$ are canonically 
conjugate
\begin{displaymath} 
\{\tilde{E}_{a}^{m}(x),A_{nb}(x')\}  = 
-i\, \delta_{ab}\delta_{n}^{m}\delta^{(3)}(x-x')
\end{displaymath}
and subject to the first class constraints
\begin{eqnarray}
\mathcal{H}\,\, & :=& \varepsilon^{abc} 
F^{a}_{mn}\tilde{E}^{mb}\tilde{E}^{nc}\approx 0 \label{As3b}\\
\mathcal{C}_{m} & :=& F^{a}_{mn}\tilde{E}^{na}\approx 0\label{As3c}\\
\mathcal{G}^{a} & :=& D_{m}\tilde{E}^{ma}\approx 0,\label{As3d}
\end{eqnarray} 
where $D_{m}$ and $F^{a}_{mn}$ denote the covariant derivative 
with respect to the connection
$A_{nb}$ and the corresponding field strength, respectively.

Again, one chooses adapted coordinates such that the Killing vectors 
are given by 
$\frac{\partial}{\partial y}$ and $\frac{\partial}{\partial z} $.
Imposing various further gauge fixings and solving the resulting second
class constraints, one eventually arrives at a reduced phase space consisting
of the canonical pairs $(\tilde{E}_{1}^{x}, A_{x1})$ and 
$(\tilde{E}_{\bar{a}}^{\bar{m}}, A_{\bar{n}\bar{b}})$,   
($\bar{m}, \bar{n}, \ldots =y,z$ and $\bar{a}, \bar{b}, \ldots =2,3$),
which are subject to three surviving constraints, one of them being an 
$SO(2)$-remnant of the $SO(3)$-Gauss law constraint (\ref{As3d}).
The non-trivial dynamics is carried by    $\tilde{E}_{\bar{a}}^{\bar{m}}$
and $A_{\bar{m}\bar{a}}$.

Defining the $SO(2)$-invariant contractions\footnote{For later convenience, 
our prefactors are 
slightly different from those of ref \cite{MZ}}
\begin{eqnarray}
K_{\bar{m}}^{\bar{n}}&:=&-i\, 
A_{\bar{m}\bar{a}}\tilde{E}_{\bar{a}}^{\bar{n}}\label{As10a}\\
J_{\bar{m}}^{\bar{n}}&:=& 
-\varepsilon^{\bar{a}\bar{b}} A_{\bar{m}\bar{a}}
\tilde{E}_{\bar{b}}^{\bar{n}},\label{As10b}
\end{eqnarray}
with $\varepsilon^{\bar{a}\bar{b}}=-\varepsilon^{\bar{b}\bar{a}},\quad 
\varepsilon^{23}=+1$, the traceless $(2\times 2)$-matrices
\begin{eqnarray}
A_{0} & = & (A_{0})_{\bar{n}\bar{m}} := K_{\bar{m}}^{\bar{n}}- 
\frac{1}{2} K_{\bar{p}}^{\bar{p}}\delta_{\bar{m}}^{\bar{n}}\label{As16a}\\
A_{1} & = & (A_{1})_{\bar{n}\bar{m}} := J_{\bar{m}}^{\bar{n}}- 
\frac{1}{2} J_{\bar{p}}^{\bar{p}}\delta_{\bar{m}}^{\bar{n}}
\label{As16b}
\end{eqnarray}
are found to obey
\begin{eqnarray}
\partial_{0}A_{0} -\partial_{1}A_{1} & = & 0\label{sum6a}\\
\partial_{0}A_{1} -\partial_{1}A_{0} + \frac{1}{\rho}[A_{0},A_{1}] & = & 0,
\label{sum6b}
\end{eqnarray}
which  has a striking similarity with (\ref{sum1}) and (\ref{sum2}).
As opposed to the currents $J_{\mu}$, however, the Poisson brackets 
between the matrices $A_{\mu}$ are completely ultralocal (see section 4). 

The link between the variables $J_{\mu}$ and $A_{\mu}$ was found
in ref \cite{MZ}:
\begin{eqnarray}
A_{0} & = & \frac{1}{2} J_{0}     + \frac{i}{2}\varepsilon \partial_{x}(\rho g)
\nonumber\label{sum10a}\\
A_{1} & = & \frac{1}{2} J_{1}     + \frac{i}{2}\varepsilon 
\partial_{t}(\rho g),
\label{sum10b}
\end{eqnarray} 
where
\begin{equation}\nonumber\label{sum3d}
\varepsilon = \left(\begin{array}{cc}
0 & 1\\
-1 & 0
\end{array}\right).
\end{equation} 
Using this relation, it can be verified that the field equations
(\ref{sum6a}), (\ref{sum6b}) are equivalent to (\ref{sum1}), (\ref{sum2})
and that the non-ultralocal terms in the Poisson brackets cancel 
for the linear combinations (\ref{sum10b}) due to some special properties
of the involved $(2\times 2)$-matrices.

Having a closer look at the linear combinations (\ref{sum10b}), however,
also raises some questions. Whereas the first term $J_{\mu}$ is
clearly Lie algebra (ie.$\mathfrak{sl}(2,\R)$)-valued (cf. eq. (\ref{Jg})), 
the second term looks a bit odd. It consists of
the derivative of a group element $g$ multiplied by a matrix $\varepsilon$,
which in general does not have any abstract Lie algebraic meaning. It just
happens that $g$ is symmetric and  $\varepsilon$ is antisymmetric,
so that the tracelessness of their products allow their interpretation
as  $\mathfrak{sl}(2,\R)$-valued quantities. At first sight,   this seems
to make it unlikely that similar variables can be constructed for 
coset spaces other than $SL(2,\R)/SO(2)$ and that the $A_{\mu}$
should perhaps more be considered as a mathematical curiosity
 of $SL(2,\R)/SO(2)$ 
without a fundamental meaning. 

The following
two sections, however, will show that this is not true and 
that the variables $A_{\mu}$ of (\ref{sum10b}) can indeed be generalized
to a large number of coset spaces. We will not attempt to construct these
analogs via dimensional reduction of matter coupled gravity 
models in terms of their corresponding $4d$  Ashtekar formulation. 
Instead, we will
work entirely in two dimensions using the relation (\ref{sum10b}) as a model.
This approach has the advantage that we can make use of the
powerful  group theoretical
structure underlying the $2d$ coset space nonlinear $\sigma$-models whose 
explicit
parametrization in terms of their $4d$ ancestors is sometimes quite intricate
\cite{BGM,BM}.

\section{The conventional formulation for arbitrary symmetric spaces}
\setcounter{equation}{0}
In this section, we 
briefly recapitulate the standard formulation of more general
$2d$  nonlinear coset space $\sigma$-models as they typically arise in 
dimensional reductions of higher dimensional (super)gravity theories.
This  generalizes  (the metric version of) the construction for 
$SL(2,\R)/SO(2)$ 
given in the previous section.

Let $G$ be a simple Lie group with involutive automorphism $\tau$ 
($\tau^{2}=1$, $\tau\neq 1$) such that $H=\{g\in G: \tau(g)=g\}$
is the maximal compact subgroup of $G$ and the coset space $G/H$ is a 
noncompact Riemannian symmetric space \cite{SH}.

The involution $\tau$ induces a decomposition of the underlying Lie algebra 
$\g$ of $G$  
\begin{displaymath}
\g=\h\oplus \ka,  
\end{displaymath}
where $\h:=\{\xi\in\g:\tau(\xi)=\xi\}$ is the Lie algebra of $H$ and 
$\ka:=\{ \xi \in \g:\tau(\xi)=-\xi\}$ coincides with the orthogonal complement 
of $\h$ 
with respect to the Cartan-Killing form of $\g$. The generators of $\g$, $\ka$
and $\h$ will be denoted by $T_{M}$, $T_{m}$ and $T_{a}$, respectively.

Let $\Sigma$ be a 2-dimensional Lorentzian manifold with local coordinates $
(x^{\mu})=(x^{0},x^{1})=(t,x)$. 
$G/H$-valued fields on $\Sigma$ can be parametrized by $G$-valued functions 
\dm
\V:\Sigma\rightarrow  G
\dn
that are subject to the local gauge freedom of left $H$-multiplication
(in addition to the global action of $G$ on $G/H$):
\eq\label{trafo}
\V(x^{\mu})\rightarrow h(x^{\mu})\V(x^{\mu}) g^{-1},
\qquad h\in H, \quad g\in G.
\en
The `vielbein' $\V$ is the proper generalization of the `unimodular zweibein' 
$e_{\bar{m}}^{\bar{a}}$ introduced in eq. (\ref{ge}) for
the case $G/H=SL(2,\R)/SO(2)$. 
The Lie algebra valued currents $\partial_{\mu}\V\cdot \V^{-1}$
decompose as
\begin{equation}\label{defQP}
\partial_{\mu}\V\cdot \V^{-1}=Q_{\mu} + P_{\mu}, \qquad Q_{\mu}\in \h, \quad  
P_{\mu}\in \ka
\end{equation}
with the transformation laws (induced by (\ref{trafo}))
\eqns
Q_{\mu}&\rightarrow& hQ_{\mu}h^{-1}+\partial_{\mu}h\cdot h^{-1}\\
P_{\mu}&\rightarrow& hP_{\mu}h^{-1}.
\enns
Instead of the redundant parametrization $\V$, one can also use the 
alternative variable
\eq
 M:=\tau(\V^{-1})\V,
\en
which is manifestly $H$-invariant (and $G$-
covariant). For $G/H=SL(2,\R)/SO(2)$, $\tau$ is given by 
$\tau(\V)=(\V^{T})^{-1}$,
ie. $M$ generalizes the (unimodular) metric block 
$g_{\bar{m}\bar{n}}$ (cf. (\ref{ge})).
 The
corresponding currents
\eq
M^{-1}\partial_{\mu}M=2 \V^{-1}P_{\mu}\V
\en
transform under (\ref{trafo}) according to
\begin{equation}\label{trafoMdM}
M^{-1}\partial_{\mu}M\rightarrow g^{-1}(M^{-1}\partial_{\mu}M)g.
\end{equation}

\subsection{The field equations}
Like the matrix $e_{\bar{m}}^{\bar{a}}$ (resp. $g_{\bar{m}\bar{n}}$) in 
Section \ref{motivation}, $\V$ (resp.  $M$) will from now on be
considered as a matrix-valued function on $\Sigma$, based on some 
faithful representation of $G$. 

Dimensional reduction of higher dimensional 
(super) gravity theories
to two dimensions leads to nonlinear $G/H$ coset space $\sigma$-models 
coupled to $2d$ gravity and a dilatonic 
scalar field (plus possible fermionic fields). 
The dynamically nontrivial part  of 
(the bosonic sector of) these $2d$ field theories is encoded
in the kinetic energy term for the coset fields, which is of the following 
generic form\footnote{
As in (\ref{sum0}), we assume (at least locally) 
world-sheet coordinates in which the world-sheet metric differs 
from the flat $2d$ Minkowski metric only by a conformal factor (which 
drops out in
the part of the Lagrangian we are considering here)}
\eqn\label{Lagrange}
\mathcal{L}&=&\frac{1}{8}\rho \textrm{tr}{[}M^{-1}\partial_{\mu} M M^{-1}
\partial^{\mu} M {]}\\ \label{Lagrange1}
&=&\frac{1}{2}\rho \textrm{tr}{[}P_{\mu}P^{\mu}{]}\label{Lagrange2},
\enn
where $\eta^{\mu \nu}=\textrm{diag}(-1,+1)$ is used to 
raise and lower the $2d$ worldsheet indices, and $\rho$ is the dilaton 
field solving the
free wave equation $\Box \rho=0$.

The corresponding field equations for  $M(x^{\mu})$ read

\eq\label{eomM}
\partial^{\mu}J_{\mu}=0,
\en
where the currents $J_{\mu}$ are defined by
\eq\label{defJ}
J_{\mu}:=\rho M^{-1}\partial_{\mu}M
\en
and therefore obey the following integrability condition
\eq\label{icM}
\partial_{0}J_{1}-\partial_{1}J_{0} + \frac{1}{\rho}{[}J_{0},J_{1}{]}
-\frac{\partial_{0}\rho}{\rho}J_{1}+\frac{\partial_{1}\rho}{\rho}J_{0}=0.
\en
In terms of the variables $P_{\mu}$ and $Q_{\mu}$, the equation of motion
becomes
\eq\label{eomV}
D^{\mu}(\rho P_{\mu})=0
\en
with the H-covariant derivative 
\dm
D_{\mu} P_{\nu}:= \partial_{\mu}P_{\nu}-{[}Q_{\mu},P_{\nu}{]},
\dn
whereas the definition (\ref{defQP}) of $Q_{\mu}$ and $P_{\nu}$ entails the 
integrability conditions
\eqn
D_{0} P_{1}-D_{1}P_{0}=0\label{icV}\\
\partial_{0}Q_{1}-\partial_{1}Q_{0}+{[}P_{1},P_{0}{]} + {[}Q_{1},Q_{0}{]}=0.
\label{icV2}
\enn

\subsection{The Poisson structure}
Extracting the Poisson brackets from the Lagrangian (\ref{Lagrange2})
requires some care due to the coset properties of the field variables.
We will simply quote the results and refer to ref. \cite{KS3} for a detailed 
description of this procedure.

Let us first introduce some notation. For any $(n\times n)$-matrices 
$A=A_{\alpha\beta}$ and $B=B_{\gamma\delta}$ we define

\dm
\stackrel{\:1}{A}:=A\otimes \mathbf{1}, \qquad \stackrel{2}{A}:= \mathbf{1}
 \otimes A
\dn
and similarly for $B$\footnote{In components, 
  $(A\otimes\mathbf{1})_{\alpha\beta,\gamma\delta}\equiv A_{\alpha\beta}
\delta_{\gamma\delta}$ and $(\mathbf{1}\otimes A)_{\alpha\beta,\gamma\delta}
\equiv A_{\gamma\delta} 
\delta_{\alpha\beta}$ etc.}.

For the Poisson brackets between the components of $A$ and $B$, we
introduce \cite{FT}
\dm
\{\stackrel{\:1}{A},\stackrel{2}{B} \}_{\alpha\beta,\gamma\delta}
:= \{A_{\alpha\beta},B_{\gamma\delta}\}.
\dn

Finally, $\beta^{MN}$ denotes the inverse of 
\begin{displaymath}
\beta_{MN}:= \textrm{tr}(T_{M}T_{N}).
\end{displaymath}

The  non-vanishing Poisson brackets for the set of variables 
$(P_{\mu},Q_{\mu})$ can then be concisely written as
\eqn
\{ {\stackrel{1}{P}}_{0}(x), {\stackrel{2}{P}}_{1}(y)\}
&=& -\frac{1}{\rho(x)}\left({[}\Omega_{\ka},
\stackrel{1}{Q_{1}}{]}\delta (x-y) +\Omega_{\ka}\partial_{x}
\delta(x-y)\right)\label{PB1}\\ 
\{ {\stackrel{1}{P}}_{0}(x), {\stackrel{2}{Q}}_{1}(y)\}&=& 
-\frac{1}{\rho}{[}\Omega_{\h},
\stackrel{1}{P_{1}}{]}\delta (x-y) \label{PB2},
\enn
where $\Omega_{\ka}:=\beta^{mn}T_{m}\otimes T_{n}$ and
$\Omega_{\h}:=\beta^{ab}T_{a}\otimes T_{b}$.
(Note that  the arguments of functions multiplying derivatives of $\delta(x-y)$
have to be treated with some care.)

Remembering $\partial_{1}\V \V^{-1} = 
Q_{1} + P_{1}$, eqs. (\ref{PB1}) and (\ref{PB2}) imply
\eqn
\{ {\stackrel{1}{P}}_{0}(x), \stackrel{2}{\V}(y)\} 
&=& \frac{1}{\rho}\Omega_{\ka}\stackrel{2}{\V}\delta
(x-y).
\enn
Using these Poisson brackets, one can then also determine 
the Poisson brackets between the $J_{\mu}$:
\eqn
\{ \stackrel{1}{J}_{0}(x), \stackrel{2}{J}_{0}(y)\} & = & 
2 {[}\Omega_{\g}, 
\stackrel{1}{J}_{0} {]} \delta(x-y)\\
\{ \stackrel{1}{J}_{1}(x), \stackrel{2}{J}_{0}(y)\} & = & 2 
{[}\Omega_{\g}, 
\stackrel{1}{J}_{1} {]} \delta(x-y)\nonumber\\
 & & -4\left(\rho\stackrel{1\quad}{\V^{-1}}
\stackrel{2\quad}{\V^{-1}}\Omega_{\ka} \stackrel{1}{\V}
\stackrel{2}{\V}\right)(x)\cdot\partial_{x}\delta (x-y)\label{PBJ2}\\
\{ \stackrel{1}{J}_{1}(x), \stackrel{2}{J}_{1}(y)\} &=& 0,
\enn
where $\Omega_{\g}:= \beta^{MN}T_{M}\otimes T_{N}=\Omega_{\ka}+\Omega_{\h}$
denotes the Casimir element of $\g$.
Note that for both sets of variables, $(P_{\mu}, Q_{\mu})$ as well as 
$(J_{\mu})$, 
the Poisson brackets involve non-ultralocal terms (i.e. terms containing
derivatives of $\delta(x-y)$).

\section{Ashtekar-type  variables for Hermitian symmetric coset spaces}
\setcounter{equation}{0}
As has been explained in section \ref{motivation}, the 
nonlinear $\sigma$-model based on the coset space 
$SL(2,\R)/SO(2)$ admits the construction of an alternative set of variables,
which have a natural embedding into the Ashtekar formulation
of $4d$ general relativity.  In view of the attractive features of these
variables (slightly simplified equations of motion plus ultralocal Poisson 
brackets), it is natural to ask whether they can be generalized to other 
symmetric spaces $G/H$. 

A first hint comes from the coset space $SL(2,\R)/SO(2)$ itself. It
is a so-called Hermitian symmetric space,
ie. a symmetric space that admits a complex structure (see eg. \cite{SH}
for a precise definition). Since the variables (\ref{sum10b}) are obviously 
based on a 
complexification, 
one might  suspect that Hermitian symmetric spaces could provide natural 
candidates
for a generalization of the relation (\ref{sum10b}).
We will now show that this is indeed the case. 

The Hermiticity of a Hermitian symmetric space $G/H$ is reflected
in its peculiar group theoretical structure:  All Hermitian
symmetric spaces $G/H$ are  of the form
\begin{displaymath}
G/H=G/(H'\times U(1))
\end{displaymath}
with some compact group $H'$. 

It is the (properly normalized) $U(1)$-generator $u$ which induces the 
complex structure on the coset space $G/H$:

\begin{equation}\label{uk}
{[}u,{[}u,k{]}{]}=-k, \textrm{ for all } k\in\ka.
\end{equation}

Having a closer look at the mechanism that led to a cancellation
of the non-ultralocal terms in the Poisson brackets for the combinations 
(\ref{sum10b})
in the $SL(2,\R)/SO(2)$-model, one finds that it is precisely 
the property  (\ref{uk}) (plus the trivial identity
$[u,\h]=0$) which is needed
for ultralocality. Thus, Hermitian symmetric spaces provide exactly 
the right amount of additional structure that allows the extension
of the construction (\ref{sum10b}) beyond $SL(2,\R)/SO(2)$.

Let us now become more explicit. Consider 
\eqn
A_{0}&:=& \rho \V^{-1}\left[P_{0}+i{[}u,P_{1}{]} + i 
\frac{\partial_{x}\rho}{\rho}u\right]\V\label{defA}\\
 & \equiv &\frac{1}{2}J_{0}+i\partial_{x}(\rho\V^{-1}u\V)\label{defA1}\\
A_{1}&:=&\rho \V^{-1}\left[ P_{1}+i{[}u,P_{0}{]} + i 
\frac{\partial_{t}\rho}{\rho}u\right] \V\\
 & \equiv &\frac{1}{2}J_{1}+i\partial_{t}(\rho\V^{-1}u\V).\label{defA2}
\enn
These variables are manifestly
$\g^{\mathbb{C}}$-valued, where $\g^{\mathbb{C}}$
denotes the complexification of the Lie algebra
$\g$. Remembering $[u,\h]=0$, it is also easy to see that the $A_{\mu}$
are $H$-gauge invariant (cf. (\ref{trafo}), (\ref{trafoMdM})),  
which generalizes the $SO(2)$-invariance of the $A_{\mu}$ of section 
\ref{motivation}.

As in the case $SL(2,\R)/SO(2)$, the advantage of these complexified 
potentials is twofold. 
First, consider the equations of motion. A straightforward calculation
reveals (cf. (\ref{eomV}) and (\ref{icV}))
\eqn
\partial^{\mu}A_{\mu} & = & \V^{-1}\left(
D^{\mu}(\rho P_{\mu})+i\rho{[}u,\varepsilon^{\mu\nu}D_{\mu}P_{\nu}{]}
\right)\V\nn\\
&=&0\nn\\
\partial_{0}A_{1}-\partial_{1}A_{0} +\frac{1}{\rho}{[}A_{0},A_{1}{]}&=&
 -\V^{-1}(\rho\varepsilon^{\mu\nu}D_{\mu}P_{\nu}+i
{[}u,
D^{\mu}(\rho P_{\mu}){]}\nn\\
& &\qquad\qquad\qquad\qquad\qquad+i\Box\rho u)\V\nn\\
&=&0
\enn
with  
$\varepsilon^{\mu\nu}=-\varepsilon^{\nu\mu}$, $\varepsilon^{10}=+1$.
Obviously, the equations of motion for 
$A_{\mu}$ are again of a similar
but somewhat simpler form  compared to the equations (\ref{eomM}) and
(\ref{icM}) for the currents $J_{\mu}$, although both sets of equations
are completely equivalent. In particular, terms like the ones
in (\ref{icM}) involving the logarithmic derivatives of $\rho$ are absent.
It is \emph{precisely} due to  these latter terms that the spectral parameter
of the linear system encoding 
(\ref{eomM}) and (\ref{icM}) has to have an irrational  $x^{\mu}$-dependence
\cite{DM,BZ},
as will become clear in the next section.
 
Another simplification occurs with respect to the Poisson brackets.
Using the techniques of the previous section, one finds, after some algebra,
\eqn
\{ {\stackrel{1}{A}}_{0}(x), {\stackrel{2}{A}}_{0}(y)\} 
&=&{[}\Omega_{\g},\stackrel{1}{A_{0}}{]}
\delta(x-y)\nn\\
\{ {\stackrel{1}{A}}_{1}(x), {\stackrel{2}{A}}_{0}(y)\} 
&=&{[}\Omega_{\g},\stackrel{1}{A_{1}}{]}
\delta(x-y)\nn\\
\{ {\stackrel{1}{A}}_{1}(x), {\stackrel{2}{A}}_{1}(y)\} 
&=&{[}\Omega_{\g},\stackrel{1}{A_{0}}{]}
\delta(x-y),\label{PBA}
\enn
which is completely ultralocal as opposed to the Poisson brackets between the 
$P_{\mu}$ or the $J_{\mu}$. The (quite non-trivial) cancellation of
the non-ultralocal terms is due to a subtle and well-balanced interplay
between the different terms appearing in the definition of the $A_{\mu}$
(\ref{defA})-(\ref{defA2}) and the specific properties of the generator $u$.

This interplay  does not work anymore for the
  Poisson brackets
between the $A_{\mu}$ and their complex conjugates, where the non-ultralocal 
terms can no longer  
be bypassed. However, the Poisson bracket between $A_{\mu}$
and ${\bar{A}}_{\mu}$ is not needed in  a 
canonical formulation along the lines of \cite{AA}, which can be seen  from
  the original definition of the $A_{\mu}$
for $SL(2,\R)/SO(2)$ in terms of the $4d$ Ashtekar variables 
(eqs. (\ref{As10a}) - (\ref{As16b})): Obviously, 
the complex conjugate
of $A_{\mu}$ involves the complex conjugate ($4d$) Ashtekar connection,
 whose Poisson bracket with the original ($4d$) Ashtekar connection is\\ 
a) already non-ultralocal in four dimensions and\\
b) completely irrelevant for this  formulation of general relativity 
\cite{AA}.

We conclude this section with some more  ``phenomenological'' remarks.
In order to get an impression of what we are talking about, let us 
first give a 
complete list of the possible non-compact Hermitian symmetric spaces \cite{SH}
(The corresponding compact versions would also allow the construction
of $A_{\mu}$-like quantities, but compact coset spaces cannot occur  in 
dimensional reductions, as they cannot contain the non-compact
space $SL(2,\R)/SO(2)$ from pure $4d$ gravity as a subspace.)  
\begin{itemize}
\item $Sp(2n,\R)/U(n)$
\item $SO^{*}(2n)/U(n)$
\item $SU(n,m)/S(U(n)\times U(m))$
\item $SO(n,2)/SO(n)\times SO(2)$
\item $E_{6(-14)}/SO(10)\times SO(2)$
\item $E_{7(-25)}/E_{6}\times U(1)$
\end{itemize}

As for the higher dimensional origin of these theories, an interesting 
observation was made in a recent paper by Cremmer, Julia, L\"{u}
and Pope \cite{CJLP}. There it was pointed out that these $2d$ theories
with Hermitian symmetric coset spaces $G/H$ all have their `oxidation endpoint'
in four dimensions, ie. the highest possible dimension for a theory
that, upon dimensional reduction, leads to the above Hermitian symmetric space
nonlinear $\sigma$-models is four (generic coset space models can 
usually be `oxidized' to much higher dimensions $d\leq 11$).

The corresponding four-dimensional theories that lead to Hermitian symmetric 
 spaces in two dimensions are well-known
 and  can be 
found in
\cite{BGM, BM}. 
Some interesting special cases are:

\begin{itemize}
\item $Sp(2,\R)/U(1)\cong SL(2,\R)/SO(2)$\\
As seen in section 2,   this coset space arises in the two Killing vector 
reduction of pure $4d$ 
general relativity.
\item $SU(2,1)/S(U(2)\times U(1))$ and the higher dimensional analogs 
$SU(n,1)/S(U(n)\times~U(1))$\\
They result from $4d$ Maxwell-Einstein gravity, respectively its 
generalizations with $(n-1)$ vector fields.
\item $SO(3,2)/SO(3)\times SO(2)\cong Sp(4,\R)/U(2)$\\
This coset space occurs in the dimensional reduction of $4d$ 
Maxwell-Einstein-dilaton-axion theory, where
the two  \emph{four-dimensional} scalar fields parametrize $SO(2,1)/SO(2)\cong 
SL(2,\R)/SO(2)$
\item $SO(8,2)/SO(8)\times SO(2)$\\
The corresponding $4d$ ancestor of this model is  (the bosonic sector of) 
$d=4$, $\mathcal{N}=4$ supergravity.

\end{itemize}

We finally note that if, like in the fundamental representations of
$Sp(2n,\R)$, $SO^{*}(2n)$ and $SU(m,m)$, $u$ is represented by an 
invertible
 matrix with $u^{-1}ku=-k$ for all $k\in \ka$, the involution $\tau$ is
  given by
\dm
\tau(\xi)=u^{-1}\xi u , \quad \xi \in \g.
\dn
In such a case, the $A_{\mu}$ can be written as
\eqn\label{AM}
A_{0}&=& \frac{1}{2}J_{0}+iu\cdot\partial_{x}(\rho M)\nn\\
A_{1}&=& \frac{1}{2}J_{1}+iu\cdot\partial_{t}(\rho M).
\enn
In the fundamental representation of $SU(m,m)$, $u$ is given by
\dm
u=\frac{i}{2}\left(\begin{array}{cc}\mathbf{1}_{m}&0\\
                               0& -\mathbf{1}_{m}
\end{array}\right),
\dn
whereas for $Sp(2n,\R)$ and $SO^{*}(2n)$ one has
\dm
u=\frac{1}{2}\left(\begin{array}{cc}0&\mathbf{1}_{n}\\
                               -\mathbf{1}_{n}&0
\end{array}\right)
\dn
so that eqs.\ (\ref{AM}) reduce to the form (\ref{sum10b}) for 
the two Killing vector reduction of pure $4d$ general relativity, 
where $G/H=SL(2,\R)/SO(2)\cong Sp(2,\R)/U(1)$.

To sum up, we have  obviously found the direct generalization of
the Ashtekar-type variables (\ref{sum10b}) for arbitrary Hermitian 
symmetric spaces $G/H$.
Although all the $4d$ ancestors of these models are known, it is an 
open question at this point whether the $A_{\mu}$ for $G/H\neq SL(2,\R)/SO(2)$
also have their natural origin
in a corresponding $4d$ Ashtekar formulation. On the other hand, from a purely 
practical point of view, this is not very important.
One can simply try to work with these variables without really having to know
 where they might come from. Adopting this attitude for the moment, we will
now have a closer look at the integrability of the field equations.

\section{The Lax pair for the complexified potentials}\setcounter{equation}{0}
It is a well-known fact \cite{DM,BZ} that the field equations (\ref{eomM}),
(\ref{icM}) 
for the currents $J_{\mu}$, respectively (\ref{eomV})-(\ref{icV2}) 
for $Q_{\mu}$ and 
$P_{\mu}$, are completely integrable in the sense that they can be written as 
the compatibility condition of a system of linear differential equations
(`Lax pair').  

These linear systems can be cast into a very compact form if they
are written in terms of   light cone coordinates $x^{\pm}:=(x^{0}\pm x^{1})$
with $\partial_{\pm}:=\frac{1}{2}(\partial_{0}\pm\partial_{1})$ and
analogously $V_{\pm}:=\frac{1}{2}(V_{0}\pm V_{1})$ for any  $V_{\mu}$.
 
Let $\tilde{\rho}$ be the harmonic conjugate to
the dilaton field $\rho$
\begin{displaymath}
\partial_{\mu}\tilde{\rho}=\varepsilon_{\mu\nu}\partial^{\nu}\rho, 
\end{displaymath}
whose (local) existence is guaranteed by $\Box \rho =0$.

Consider now the function (the `variable spectral parameter')
\begin{equation}\label{gamma}
\gamma(t,x;w):=\frac{1}{\rho}(w+\tilde{\rho}-\sqrt{(w+\tilde{\rho})^2-\rho^2}),
\end{equation}
where $w$ is a constant and the implicit $(t,x)$-dependence is via $\rho$ and
$\tilde{\rho}$.

(\ref{gamma}) can be inverted
\begin{displaymath}
w=\frac{1}{2}\rho\left(\gamma+\frac{1}{\gamma}\right)-\tilde{\rho}
\end{displaymath}
and implies
\begin{displaymath}
\gamma^{-1}\partial_{\pm}\gamma=\frac{(1\mp\gamma)}{(1\pm \gamma)}\rho^{-1}
\partial_{\pm}\rho.
\end{displaymath}

This particular spacetime dependence of $\gamma$ ensures that
the compatibility condition of the linear system
\begin{equation}
\partial_{\pm}\hat{\V} {\hat{\V}}^{-1}=Q_{\pm}+\frac{1\mp \gamma}{1\pm 
\gamma}P_{\pm}
\end{equation}
for the $G$-valued function $\hat{\V}=\hat{\V}(t,x;\gamma(t,x;w))$ 
implies the field equations (\ref{eomV})-(\ref{icV2}) for the currents 
$P_{\mu}$ and $Q_{\mu}$.

Similarly, the linear system for the $G$-valued function
$\Psi=\Psi(t,x;\gamma(t,x;w))$
\begin{equation}\label{Psi}
\Psi^{-1} \partial_{\pm}\Psi=\frac{J_{\pm}}{\rho (1\pm \gamma)}
\end{equation}
can be easily verified to imply
\begin{eqnarray}
\partial^{\mu}J_{\mu}&=&0\label{dJ}\\
\partial_{0}J_{1}-\partial_{1}J_{0} + \frac{1}{\rho}{[}J_{0},J_{1}{]}
-\frac{\partial_{0}\rho}{\rho}J_{1}+\frac{\partial_{1}\rho}{\rho}J_{0}&=&
0,\label{dJ2}
\end{eqnarray}
ie. the field equations (\ref{eomM}) and (\ref{icM}) for the currents 
$J_{\mu}$.

In both linear systems,
$\gamma$ plays the r\^{o}le of a spacetime dependent spectral parameter,
whereas $w$ can be interpreted as a `hidden' constant spectral parameter. 
In order to avoid the explicit appearance of the square roots in the linear
system, the latter is usually stated in terms of  the variable spectral 
parameter $\gamma$. In \cite{KS1} it was shown that it is exactly this 
spacetime dependence of $\gamma$ which serves as a natural (classical)
regulator that removes certain ambiguities in the Poisson brackets
between  transition matrices that are caused by the non-ultralocal terms 
in the Poisson brackets (\ref{PB1}) resp. (\ref{PBJ2}). These transition 
matrices are closely related to the infinite number of conserved charges 
that generate the Geroch group on the phase space. Quantization of the
Poisson structure of these charges led to certain twisted Yangian algebras
\cite{KS3}.

It would now be extremely interesting to see whether one arrives at 
similar structures if one quantizes the system in terms of 
the generalized Ashtekar variables $A_{\mu}$. 
In order to do so, it would obviously be very convenient if one could
also make use of a linear system for the equations of motion 
\begin{eqnarray}
\partial^{\mu}A_{\mu}&=&0\label{dA}\\
\partial_{0}A_{1}-\partial_{1}A_{0} + \frac{1}{\rho}{[}A_{0},A_{1}{]}&=&0
\label{dA2},
\end{eqnarray}
for 
the complexified potentials $A_{\mu}$. In fact, one might suspect that the
simplicity of these equations should be reflected
in a simpler linear system compared to the ones for
the variables ($J_{\mu}$) or ($P_{\mu}, Q_{\mu}$). In particular, one might 
hope that the spacetime dependence of the spectral parameter might
simplify. And indeed, this is exactly what happens.
The linear system encoding (\ref{dA})-(\ref{dA2}) has the same form as
the system (\ref{Psi}) for the equations (\ref{dJ})-(\ref{dJ2}):
\begin{equation}\label{U}
U^{-1} \partial_{\pm}U=\frac{A_{\pm}}{\rho (1\pm \lambda)},
\end{equation}
but the variable spectral parameter $\lambda$ is now given by
\begin{equation}\label{lambda}
\lambda(t,x;v):=\frac{1}{\rho}(v+\tilde{\rho})
\end{equation}
with the `hidden' constant spectral parameter
\begin{displaymath}
v=\rho \lambda -\tilde{\rho}
\end{displaymath}
and the resulting differential equation
\begin{displaymath}
\partial_{\pm}\lambda=\pm(1\mp\lambda)\rho^{-1}
\partial_{\pm}\rho.
\end{displaymath}

This shows that   the square roots in the spacetime dependence of 
the variable spectral parameter $\gamma$ (eq. (\ref{gamma}))
for the $J_{\mu}$ is precisely due to the extra terms in (\ref{dJ2}) which 
involve the 
logarithmic derivatives of 
$\rho$ and which are missing in (\ref{dA2}).    
In view of the lack of these square roots,   there is   no 
reason anymore to `hide' 
the constant spectral parameter $v$, and one can write the linear system
(\ref{U}) entirely in terms of $v$ without introducing square roots:
 \begin{equation}\label{Uv}
U^{-1} \partial_{\pm}U=\frac{A_{\pm}}{\rho \pm (v+\tilde{\rho})}.
\end{equation}

This meshes nicely with the ultralocality of
the Poisson brackets (\ref{PBA}) between the $A_{\mu}$, for now a 
regularization mechanism induced by a particular  spacetime dependence
in order to remove ambiguities in the Poisson
structure of the transition matrices is not necessary anymore.

In fact, it is now comparatively easy to verify that the transition matrices
\begin{eqnarray}\label{defT}
T(v)&:=&U^{-1}(t,x=-\infty;v)U(t,y=\infty;v)\\ 
    &=& \mathcal{P} \exp \left[ \int_{-\infty}^{\infty} dx 
\frac{\rho A_{1}-(v+\tilde{\rho})A_{0}}{\rho^{2} - 
(v+\tilde{\rho})^{2}} \right],\nonumber 
\end{eqnarray}
where $\mathcal{P}$ denotes the path-ordered exponential,
have the Poisson algebra
\begin{equation}
\{\stackrel{1}{T}(v), \stackrel{2}{T}(w)\}=\frac{1}{v-w}\left[\stackrel{1}{T}
(v)\stackrel{2}{T}(w),
\Omega_{\g}\right].
\end{equation}
Formally, this Poisson algebra looks the same as the one found in the 
metric formalism \cite{KS1}, yet the respective transition matrices 
themselves might a priori have nothing to do with each other. 

The definition (\ref{defT}) of the transition matrices and the linear system 
(\ref{Uv}) imply that the $T(v)$ are time-independent for sufficiently rapidly 
decreasing  boundary conditions and therefore comprise an infinite set
of conserved non-local charges parametrized by a parameter $v$. 

A formal expansion 
\begin{displaymath}
T(v)=\mathbf{1}+ \frac{T_{1}}{v}+\frac{T_{2}}{v^{2}}+\ldots
\end{displaymath}
yields for the first three non-trivial charges
\begin{eqnarray}
T_{1}&=&\int_{-\infty}^{\infty}dx A_{0}\\
T_{2}&=& \int_{-\infty}^{\infty}dx(-\rho A_{1}-\tilde{\rho} A_{0})+
\int_{-\infty}^{\infty}dx A_{0}\int_{x}^{\infty}dyA_{0}\label{T2}\\
T_{3}&=& \int_{-\infty}^{\infty}dx \left[2\rho\tilde{\rho}A_{1}+
(\rho^{2}+{\tilde{\rho}}^{2})A_{0}\right] +
\int_{-\infty}^{\infty}dx A_{0}\int_{x}^{\infty}dy (-\rho A_{1}-\tilde{\rho}
A_{0})\nonumber\\
&+&\int_{-\infty}^{\infty}dx(-\rho A_{1}-\tilde{\rho} A_{0})
\int_{x}^{\infty}dy A_{0} +\int_{-\infty}^{\infty}dx A_{0}\int_{x}^{\infty}dy 
A_{0}\int_{y}^{\infty}dz 
A_{0}.
\end{eqnarray}
(Using the equations of motion (\ref{dA}) - (\ref{dA2}), the time 
independence of these charges can also  be verified directly.)

It is instructive to compare these charges with the analogous charges 
found in the metric approach \cite{KS1}. The first charge $T_{1}$ is the
same as the corresponding charge in \cite{KS1}, since the imaginary part of 
$A_{0}$ is a total derivative and the real part is just $J_{0}/2$ (see eq.
(\ref{defA1})). Interestingly enough, the second charge, $T_{2}$, also
reproduces the corresponding charge in  \cite{KS1}, when the
definition (\ref{defA1})-(\ref{defA2})  is inserted into (\ref{T2})
and some partial  integrations are performed. Since these manipulations,
in particular, the above series expansion of $T(v)$, are rather
formal and might possibly require some more care, 
we leave a further comparison of  these two sets of charges,
or rather the corresponding transition matrices,
as an interesting problem for the future.

\section{Conclusion and open problems}\setcounter{equation}{0}

In this article we have given a reformulation of a certain class of 
dimensionally reduced (super)gravity theories. These theories are 
described by $2d$ dilaton 
gravity 
coupled $G/H$ nonlinear $\sigma$-models for which $G/H=G/(H'\times U(1))$ 
is a Hermitian symmetric space. This construction was motivated by the
formulation of the two Killing vector reduction of 
pure $4d$ gravity in terms of the Ashtekar variables, where the coset space
is $SL(2,\R)/SO(2)$. In this sense, our construction can be understood as a 
direct (ie.  two-dimensional)
generalization of the dimensionally reduced
Ashtekar formulation to arbitrary Hermitian symmetric 
 coset spaces. At this point, however, this is, strictly speaking,
 still an  analogy, and it
would be very interesting to see whether the variables constructed
in this paper really have a natural embedding into the corresponding
$4d$ Ashtekar formulations also for $G/H\neq SL(2,\R)/SO(2)$.

The theories with Hermitian symmetric coset spaces in $2d$ have
recently been found to be very special also in that they all have their 
`oxidation 
endpoints' in $4d$. Since the Ashtekar formulation of general relativity
is essentially tied to four dimensions as well, there seems to be  
a certain consistency with our results. 
Having a closer look at the 
simplest cases like eg. $4d$ Maxwell-Einstein gravity might 
perhaps help to understand these connections a little bit better.

For purely practical purposes, on the other hand, the exact  knowledge 
of a potential
$4d$ origin is not very relevant. 
One can simply start with the 
two-dimensional theories, perform the transformation to the new 
set of variables and then try to exploit some of their attractive properties.

These attractive properties are twofold:\\
1) The Poisson brackets are completely ultralocal. \\
2) The equations of motion are of a slightly simpler form. 
As we have shown, a linear system can also be constructed for these 
equations of motion, and the simpler form of the field equations
is reflected in a   simpler (in fact, easily removable) spacetime 
dependence of the 
spectral parameter. The mere existence of 
this linear system enables one to apply  
the same techniques that have been  used in  the metric 
formulation
\cite{KS1,KS2,KS3}.

Making this idea explicit, we exploited  the technical advantages
of the new variables
and calculated the Poisson brackets between the corresponding
transition matrices. In the metric 
picture, an analogous calculation had to deal with an intricate mechanism
that removed the ambiguities due to the non-ultralocal Poisson brackets.
In terms of the Ashtekar-type variables, however,  this ambiguity was 
absent right from 
the beginning.

If one can  find a nice way to implement the
reality conditions for the $A_{\mu}$, one might now be able
to directly probe the \emph{quantum} equivalence of the metric and the 
self-dual connection approach to quantum gravity within a
nontrivial toy model that still exhibits many of the  most important 
properties of full $4d$ general relativity.\\
\vspace{1cm}\\
\textbf{Acknowledgements}\\
We  would like to thank Abhay Ashtekar,  Domenico Giulini, Dieter Maison
and Henning Samtleben for helpful comments and discussions.


\begin{thebibliography}{99}


\bibitem{RG} R. Geroch, J.Math.Phys. \textbf{13} (1972) 394
\bibitem{BJ} B. Julia, \emph{Group disintegrations} in \emph{Superspace
and Supergravity}, S. Hawking and M. Rocek (eds.), Cambridge University 
Press, Cambridge (1980)
\bibitem{HN} H. Nicolai in \emph{Recent Aspects of Quantum Fields},
H. Mitter and H. Gausterer (eds.), Springer-Verlag, Berlin (1991)
\bibitem{DM} D. Maison, Phys. Rev. Lett. \textbf{41}, 521-522 (1978)
\bibitem{BZ} V. Belinskii and V. Zakharov, Sov. Phys. JETP \textbf{48},
985-994 (1978)
\bibitem{KS1} D. Korotkin and H. Samtleben, Class.Quant.Grav. \textbf{14}
(1997) L151; gr-qc/9611061 
\bibitem {KS2} D. Korotkin and H. Samtleben, Phys.Rev.Lett.  \textbf{80}
(1998) 14; gr-qc/9705013 
\bibitem{KS3} D. Korotkin and H. Samtleben, Nucl. Phys. \textbf{B527}, 
657-689 (1998); hep-th/9710210
\bibitem{AA} A. Ashtekar, Phys. Rev. Lett. \textbf{57} 2244 (1986);
A. Ashtekar, Phys.Rev. \textbf{D36} 1587 (1987)
\bibitem{HS} V. Husain and L. Smolin, Nucl. Phys. \textbf{B327}, 205 (1989)
\bibitem{VH1} V. Husain, Phys. Rev. \textbf{D53}, 4327-4334 (1996); 
gr-qc/9602050
\bibitem{VH2} V. Husain, Phys. Rev. \textbf{D56} (1997) 1831-1835;  
gr-qc/9703087
\bibitem{GM} G. Mena Marug\`{a}n, Phys.Rev \textbf{D56} (1997) 908;
gr-qc/9704041
\bibitem{MZ} M. Zagermann, Class. Quantum Grav. \textbf{15}, 1367-1374 (1998);
gr-qc/9710133
\bibitem{CJLP} E. Cremmer, B. Julia, H. L\"{u}, C.N. Pope, hep-th/9909099
\bibitem{BGM} P. Breitenlohner, G. Gibbons and D. Maison, Commun. Math. Phys.
\textbf{120}, 295-333 (1988)
\bibitem{BM} P. Breitenlohner and D. Maison, gr-qc/9806002    
\bibitem{SH} S. Helgason, \emph{Differential Geometry, Lie Groups and
Symmetric Spaces}, Academic Press, New York (1978)
\bibitem{FT} L. Faddeev and L. Takhtajan, \emph{Hamiltonian Methods in 
the Theory of Solitons}, Springer-Verlag, Berlin (1987)

\end{thebibliography}
\end{document}